\documentclass[runningheads]{llncs}
\usepackage[T1]{fontenc}

\usepackage{graphicx}
\usepackage{tikz}

\begin{document}

\title{A Geometry-Sensitive Quorum Sensing Algorithm for the Best-of-N Site 
Selection Problem}

\titlerunning{Geometry-Sensitive Algorithm for N-Site Selection}

\author{Grace Cai\orcidID{0000-0001-5929-4347} \and
Nancy Lynch\orcidID{0000-0003-3045-265X}}

\authorrunning{G. Cai and N. Lynch}

\institute{Computer Science and Artificial Intelligence Laboratory, MIT, Cambridge, MA, USA\\ 
\email{gracecai@mit.edu, lynch@csail.mit.edu}}
\index{Cai, Grace}
\index{Lynch, Nancy}

\maketitle             
\begin{abstract}

The house hunting behavior of the Temnothorax albipennis ant allows the colony to explore several nest choices and agree on the best one. Their behavior serves as the basis for many bio-inspired swarm models to solve the same problem. However, many of the existing site selection models in both insect colony and swarm literature test the model's accuracy and decision time only on setups where all potential site choices are equidistant from the swarm's starting location. These models do not account for the geographic challenges that result from site choices with different geometry. For example, although actual ant colonies are capable of consistently choosing a higher quality, further site instead of a lower quality, closer site, existing models are much less accurate in this scenario. Existing models are also more prone to committing to a low quality site if it is on the path between the agents' starting site and a higher quality site. We present a new model for the site selection problem and verify via simulation that is able to better handle these geographic challenges. Our results provide insight into the types of challenges site selection models face when distance is taken into account. Our work will allow swarms to be robust to more realistic situations where sites could be distributed in the environment in many different ways. 
\end{abstract}
\section{Introduction}\label{sec:intro}
\vspace{-.2cm}

Swarms of birds, bees, and ants are able to coordinate themselves to make decisions using only local interactions \cite{camazine1999house,pratt2005quorum,reynolds1987flocks}. Modelling these natural swarms has inspired many successful swarm algorithms \cite{fan2020review}. One such bio-inspired algorithm comes from the house hunting behavior of ants. Models of the ants' behavior when selecting a new nest serve as the basis for swarm algorithms which seek to select the best site out of a discrete number of candidate sites in space \cite{reina}. 
\vspace{-.03cm}

Many variations of the best-of-N site selection problem have been studied for swarms \cite{valentini2017best}. For example, when sites are of equal quality, choosing one is a symmetry-breaking problem \cite{hamann2012analysis,wessnitzer2003collective}. Situations with asymmetric site qualities and costs (where higher quality sites have a higher cost of being chosen) have also been studied -- for example, when one of two candidate sites is significantly larger than the other (making it harder for agents to detect other agents favoring the larger site, even when it is of higher quality) \cite{campo2011self}. 

However, most site selection models are mainly tested on small numbers of candidate nest sites that are equidistant from the agents' starting location (also known as the home nest) \cite{cody-adams,pratt-model}. In many applications of the site selection problem such as shelter seeking, sites will not be distributed so uniformly. 

This equidistant setup fails to capture two important geographical details that existing algorithms struggle with in making accurate decisions. Firstly, nests that are closer to the home nest are advantaged because they are more likely to be found. Even so, house hunting ants can still choose higher quality sites that are much further than lower quality, closer sites. We have found that existing site selection models often commit to the closer site even when there is a better, further option. Secondly, using sites equidistant from the home nest eliminates the possibility of some nests being in the way of others. Site selection models often trigger consensus on a new site after a certain quorum population of agents have been detected in it. If a low quality nest is on the path from the home nest to a high quality nest, agents travelling between the home nest and the high quality nest could saturate the path and detect a quorum for the lower quality nest that is in the way instead of the highest quality nest. 

This paper aims to create a new algorithm that can successfully account for a more varied range of nest distributions, allowing agents to successfully choose higher quality nests even when they have the disadvantage of being further from the agents' starting location or there are other lower quality nests in the way. The model should also perform with similar accuracy compared to existing models on the default setup with equidistant candidate nest sites. We show via simulation that incorporating a quorum threshold that decreases with site quality allows for increased accuracy compared to previous models. We also show that setups where candidate sites are in the way of each other or are of similar quality can make it harder for site selection models to produce accurate results. 

Section \ref{sec:background} describes the house hunting process of ants and overviews existing swarm models. Section \ref{sec:model} describes our model. We provide details on the implementation of our model, test accuracy and decision time in different geographic situations, and report the results in Section \ref{sec:results}. We discuss these results in Section \ref{sec:discussion}. Lastly, we suggest future work in Section \ref{sec:future}. The full simulation code can be found at \cite{Cai_Geometric_Swarm_Modelling_2022}.

\section{Background}\label{sec:background}
\subsection{Ant House Hunting}
When the \textit{T. albipennis} ants' home nest is destroyed, the colony can find and collectively move to a new, high quality nest. To do so, \textit{T. albipennis} scouts first scan the area, searching for candidate nests. When a nest is found, the scouts wait a period of time inversely proportional to the nest quality before returning to the home nest. There, they recruit others to examine the new site in a process known as forward tandem running. Tandem runs allow more ants to learn the path to a new site in case the ants decide to move there. When an ant in a candidate nest encounters others in the site at a rate surpassing a threshold rate (known as the quorum threshold), ants switch their behavior to carrying other members of the colony to the new nest. Carrying is three times faster than tandem runs and accelerates the move to the new nest \cite{pratt2005behavioral,pratt2005quorum}. 

This decision-making process allows ants to not only agree on a new 
nest, but also to choose the highest quality nest out of multiple nests in the environment. This is true even if the high quality nest is much further from the home nest than the low quality nest \cite{franks2008can,robinson2009ants}. Franks \cite{franks2008can} found that with a low quality nest 30 cm from home and a high quality nest 255cm from home, 88\% of ant colonies successfully chose the high quality nest even though it was 9 times further.

\subsection{House Hunting and Site Selection Models}
To better study the ants' behavior, models have been designed to simulate how ants change behavior throughout the house hunting process \cite{pratt-model,jiajia-model}. These models, initiated by Pratt \cite{pratt-model}, allow simulated ants to probabilistically transition through four phases -- the Exploration, Assessment, Canvassing, and Transport phases. The Exploration represents when the ants are still exploring their environment for new sites. When an ant discovers a site, it enters the Assessment phase, in which it examines the quality of the site and determine whether to accept or reject it. If the ant accepts the site, it enters the Canvassing phase, which represents the process of recruiting other ants via forward tandem runs. Finally, if a quorum is sensed, the ant enters the Transport phase, which represents the carrying behavior used to move the colony to the new site. 

These models, however, assume that when an ant transitions from the Exploration phase to the Assessment phase, it is equally likely to choose any of the candidate nest sites to assess. This assumes that any nest is equally likely to be found, which is unlikely in the real world because sites closer to the home nest are more likely to be discovered. To our knowledge, house hunting models have not tried to model situations where nests have different likelihoods of being found, as is the case when nests have different distances from the home nest \cite{jiajia-model}.  

The corresponding problem to house hunting in robot swarms is known as the \textit{$N$-site selection problem} \cite{valentini2017best}. Agents, starting at a central home site, must find and choose among $N$ candidate nest sites in the environment and move to the site with highest quality. Unlike house hunting models, which do not physically simulate ants in space, swarm models set up ants in a simulated arena and let them physically explore sites and travel between them. 

Inspired by ant modelling, \cite{cody-adams} and \cite{reina} have modeled swarm agents using four main states -- Uncommitted Latent, Uncommitted Interactive, Favoring Latent, and Favoring Interactive (with \cite{cody-adams} adding a fifth Committed state to emulate having detected a quorum). Uncommitted Latent agents remain in the home nest while Uncommitted Interactive agents explore the arena for candidate sites. Favoring Interactive agents have discovered and are favoring a certain site and recruit other agents to the site, while Favoring Latent agents remain in the new site to try and build up quorum. Agents probabilistically transition between these states based on environmental events (e.g. the discovery of a new site) and eventually end up significantly favoring a new candidate nest or committed to it. Other swarm models for N-site selection typically use a similar progression through uncommitted, favoring, and committed type phases \cite{Parker2009CooperativeDI}.

One setup where a high quality site was twice as far as a low quality one was successfully solved in \cite{reina}, but for the most part these models and their variations have mainly been tested in arenas with two candidate sites equidistant from the home nest \cite{cai,cody-adams,khurana2020quorum,reina-bee-model}. Our model aims to analyze the behavior of these models in more varied site setups and and improve upon them.
\section{Model}\label{sec:model}

We first describe our new discrete geographical model for modeling swarms. Then we discuss the individual restrictions, parameters, and agent algorithms needed for the house hunting problem specifically. 

A more formal description of our general model (described in Section 3.1) can be found in Appendix A. 

\subsection{General Model}
We assume a finite set $R$ of agents, with a state set $SR$ of potential states. Agents move on a discrete rectangular grid of size $n \times m$, formally modelled as directed graph $G = (V,E)$ with $|V| = mn$. Edges are  bidirectional, and we also include a self-loop at each vertex. Vertices are indexed as $(x, y)$, where $0 \leq x \leq n-1$, $0 \leq y \leq m-1$. Each vertex also has a state set $SV$ of potential states.

We use a discrete model so the model can be simulated in a distributed fashion on each vertex to reduce computation time. 
\vspace{-.5cm}

\subsubsection{Local Configurations:}


A \textit{local configuration} $C'(v)$ captures the contents vertex $v$. It is a triple $(sv, myagents, srmap)$, where $sv \in SV$ is the vertex state of $v$, $myagents \subseteq R$ is the set of agents at $v$, and $srmap: myagents \rightarrow SR$ assigns an agent state to each agent at $v$.


\subsubsection{Local Transitions: }
The transition of a vertex $v$ may be influenced by the local configurations of nearby vertices. We define an \textbf{influence radius} $I$, which is the same for all vertices, to mean that vertex indexed at $(x,y)$ is influenced by all valid vertices $\{(a,b) | a \in [x-I, x+I], b\in [y-I, y+I]\}$, where $a$ and $b$ are integers. 
We can use this influence radius to create a local mapping $M_v$ from local coordinates to the neighboring local configurations. For a vertex $v$ at location $(x,y)$, we produce $M_v$ such that $M_v(a,b) \rightarrow C'(w)$ where $w$ is the vertex located at $(x+a, y+b)$ and $ -I < a,b < I$. This influence radius is representative of a sensing and communication radius. Agents can use all information from vertices within the influence radius to make decisions.

We have a local transition function $\delta$, which maps all the information associated with one vertex and its influence radius at one time to new information that can be associated with the vertex at the following time.  It also produces directions of motion for all the agents at the vertex.

Formally, for a vertex $v$, $\delta$ probabilistically maps $M_v$ to a quadruple of the form $(sv_1, myagents, srmap_1, dirmap_1)$, where $sv_1 \in SV$ is the new state of the vertex, $srmap_1: myagents \rightarrow SR$ is the new agent state mapping for agents at the vertex, and $dirmap_1: myagents \rightarrow \{R,L,U,D,S\}$ gives directions of motion for agents currently at the vertex. Note that $R$, $L$, $U$, and $D$ mean right, left, up, and down respectively, and $S$ means to stay at the vertex. The local transition function $\delta$ is further broken down into two phases as follows.

\textit{Phase One:} Each agent in vertex $v$ uses the same probabilistic transition function $\alpha$, which probabilistically maps the agent's state $sr \in SR$, location $(x,y)$, and the mapping $M_v$ to a new suggested vertex state $sv'$, agent state $sr'$, and direction of motion $d \in \{R, L, U, D, S\}$. We can think of $\alpha$ as an agent state machine model. 

\textit{Phase Two:} Since agents may suggest conflicting new vertex states, a rule $L$ is used to select one final vertex state. The rule also determines for each agent whether they may transition to state $sr'$ and direction of motion $d$ or whether they must stay at the same location with original state $sr$.

\subsubsection{Probabilistic Execution: }

The system operates by probablistically transitioning all vertices $v$ for an infinite number of rounds. During each round, for each vertex $v$, we obtain the mapping $M_v$ which contains the local configurations of all vertices in its influence radius. We then apply $\delta$ to $M_v$ to transition vertex $v$ and all agents at vertex $v$. For each vertex $v$ we now have $(sv_v, myagents_v, srmap_v, dirmap_v)$ returned from $\delta$.

For each $v$, we take $dirmap_v$, which specifies the direction of motion for each agent and use it to map all agents to their new vertices. For each vertex $v$, it's new local configuration is just the new vertex state $sv_v$, the new set of agents at the vertex, and the $srmap$ mapping from agents to their new agent states. 

\subsection{House Hunting Model}
The goal of the house hunting problem is for agents to explore the grid and select the best site out of $N$ sites to migrate to collectively. We model sites as follows. 

A set $S$, $|S| = N$ of rectangular sites are located within this grid, where site $s_i$ has lower left vertex $(x_i^1, y_i^1)$ and upper right vertex $(x_i^2, y_i^2)$. Each site $s_i$ also has a quality $s_i.q \in [0,1]$. To represent these sites, we let the vertex state set be $SV = S \cup \{\emptyset \}$ for each vertex, indicating which site, if any, the vertex belongs to. Furthermore, we denote the site $s_0$ to be the \textit{home nest}. In the initial configuration, all agents start out at a random vertex in the home nest, chosen uniformly from among the 
vertices in that nest.

\subsection{Agent States and Transition Function}
\begin{figure}
    \centering
    \begin{tikzpicture}[>=latex, scale=0.8]
    \fill[fill=white]
    (0,4) node[circle,inner sep=0pt,draw, text width = 1cm, text centered](UA) {$U^A$}
 -- (4, 4) node[circle, inner sep = 0pt, draw, text width = 1cm, text centered](UN){$U^N$}
 -- (2, 2) node[circle, inner sep = 0pt, draw, text width = 1cm, text centered](FAI){$F^A_i$}
 -- (0, 0)node[circle, inner sep = 0pt, draw, text width = 1cm, text centered](FNI){$F^N_i$}
 -- (4, 0)node[circle, inner sep = 0pt, draw, text width = 1cm, text centered](FNJ){$F^N_{j \neq i}$}
 ;
\begin{scope}[every node/.style={scale=.78}]
 \path[->, sloped]
(UA) edge[bend left=5] node[above] {$P_N$} (UN)
(UN) edge[bend left=5] node[below]{$P_A$} (UA)
(UA) edge[bend left=5] node[above]{$P_{S_i} v_i (1-x)$} (FNI)
(FNI) edge[bend left=5] node[above]{$\beta$} (UA)
(FAI) edge[bend left=5] node[below] {$P_N$} (FNI)
(FNI) edge[bend left=5] node[above] {$P_A$} (FAI)
(FNI) edge[bend left=5] node[above] {$D s_j P_{ij}$} (FNJ)
(FNJ) edge[bend left=5] node[below]{$D s_i P_{ji}$}(FNI)
(UA) edge[] node[above]{$P_{S_i} v_i x$}(FAI)
;
\end{scope}
\end{tikzpicture}
    \begin{tikzpicture}[>=latex, scale=0.7]
\fill[fill=white]
    (3,0) node[circle,inner sep=0pt,draw, text width = 1cm, text centered](UA) {$U^A$}
 -- (3, 2) node[circle, inner sep = 0pt, draw, text width = 1cm, text centered](UN){$U^N$}
 -- (3,6) node[circle, inner sep = 0pt, draw, text width = 1cm, text centered](FAI){$F^N_i$}
 -- (3, 4)node[circle, inner sep = 0pt, draw, text width = 1cm, text centered](FNI){$F^A_i$}
 -- (0, 3)node[circle, inner sep = 0pt, draw, text width = 1cm, text centered](QA){$C^A_i$}
 -- (6, 3)node[circle, inner sep = 0pt, draw, text width = 1cm, text centered](QN){$C^N_i$}
 ;
 \begin{scope}[every node/.style={scale=.60}]
 \path[->, sloped]
(FAI) edge[] node[above]{$0.5(P_{Q_i}+P_{Q})$}(QN)
(FAI) edge[] node[above]{$0.5(P_{Q_i}+P_{Q})$}(QA)
(UA) edge[] node[below]{$0.5(P_{Q_i}+P_{Q})$}(QA)
(UA) edge[] node[below]{$0.5(P_{Q_i}+P_{Q})$}(QN)
(FNI) edge[] node[below]{$0.5P_{Q_i}$}(QN)
(FNI) edge[] node[below]{$0.5P_{Q_i}$}(QA)
(UN) edge[] node[above]{$0.5P_{Q_i}$}(QA)
(UN) edge[] node[above]{$0.5P_{Q_i}$}(QN)
;
\end{scope}
\end{tikzpicture}
    \caption{State model. $\{U, F, C\}$ denote preference states. The superscript $\{N, A\}$ denotes the activity state, and a subscript $i$ denotes that an agent is favoring or committed to site $i$. The transitions for Uncommitted and Favoring states are shown on the left, and transitions from Uncommitted and Favoring to Committed states are on the right.}
    \label{fig:transitions}
\end{figure}
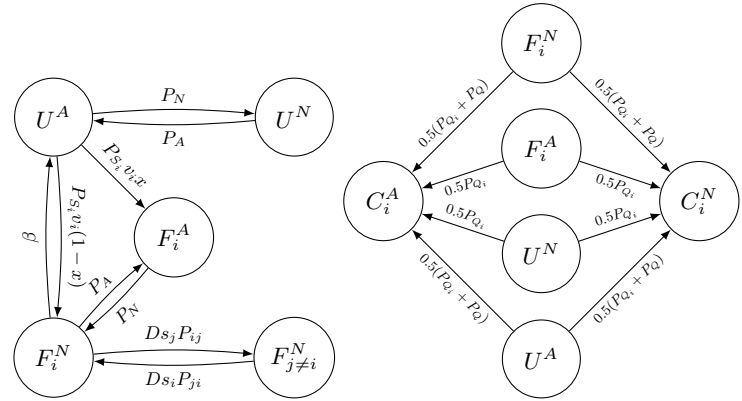

The agent state set $SR$ is best described in conjunction with the agent transition function $\alpha$. Agents can take on one of $6$ core states, each a combination of one of three preference states (Uncommitted, Favoring, Committed), and two activity states (Nest, Active). The state model can be seen in Figure \ref{fig:transitions}.

Uncommitted Nest ($U^N$) agents stay in the home nest to prevent too many agents from flooding the environment. They have a chance of transitioning to Uncommitted Active ($U^A$) agents, which try to explore the arena and discover new sites. $U^A$ agents move according to the Levy flight random walk\footnote{The Levy flight is expressed within our general model by breaking it down into individual steps, and progress along the flight is maintained in 
the agent's state. See \cite{Cai_Geometric_Swarm_Modelling_2022} for implementation details.}, which has been shown to be used by foraging ants \cite{sims2012levy}. $U^N$ agents transition to $U^A$ with probability $P_A$, and $U^A$ agents transition to $U^N$ agents with probability $P_N$. This results in an expected $x = \frac{P_A}{P_A+P_N}$ percent of uncommitted agents are active, whereas $1-x$ agents remain in the nest. Prior work \cite{reina} lets $P_N = 9P_A = L$, where $L$ is the inverse of the average site round trip time, chosen to promote sufficient mixing. This leads to $10\%$ of the agent population being active.

Uncommitted Active agents have a chance $P_{S_i}$ of discovering a new nest, which is 1 if a new nest is within influence radius and $0$ otherwise. If they discover a nest $s_i$, they explore and accept it with probability $s_i.q$ (the quality of $s_i$). They then have an $x\%$ chance of transitioning to Favoring Active, and a $(1-x)\%$ chance of transitioning to Favoring Nest. 

Favoring agents $(F^A_i, F^N_i)$ prefer the site $s_i$ that they discovered. Favoring Active $(F^A_i)$ agents remain in site $s_i$ to build quorum. Favoring Nest $(F^N_i)$ agents return to the home nest to recruit others to site $s_i$. Favoring Nest agents transition to Active with the same probability $P_A$ and Favoring Active agents transition back to Nest agents with probability $P_N$, creating the same effect where an expected $90\%$ of the favoring agent population is $F^A_i$ while the rest are $F^N_i$.

$F^N_i$ agents have a probability $\beta$ of abandoning their nest, which is 1 if the time spent without seeing other agents surpasses $t_\beta$. $F^A_i$ agents can be inhibited by other $F^A_i$ agents as follows. The chance an agent favoring nest $i$ is converted to favoring nest $j$ is $D r_j P_{ij}$, where the factor of $D$ is the probability of agents messaging each other (to prevent excessive messaging). $r_j$ is the number of agents favoring $s_j$ that have the agent within their influence radius. After an agent hears of the new site $s_j$, it visits the site to evaluate $s_j.q$ and changes its preference to $s_j$ if $s_j.q > s_i.q$. Thus, the condition $P_{ij}$ is $1$ when $s_j.q > s_i.q$ and $0$ otherwise. 

$U^A$ agents and $F^N_i$ agents can detect a quorum and commit to a site when $q$ agents in the site are within their influence radius. The quorum size scales with site value as $q = \lfloor(q_{MIN}-q_{MAX})*s_i.q + q_{MAX} \rfloor$, where $q_{MAX}$ and $q_{MIN}$ are the maximum and minimum possible quorum threshold respectively. The condition $P_Q$ is $1$ when quorum is satisfied and $0$ otherwise. Agents in any Favoring or Uncommitted state will transition to the committed state, if they encounter an agent already in quorum. The condition $P_{Q_i}$ is $1$ when another quorum agent for $s_i$ is encountered and $0$ otherwise. Furthermore, agents have an $\frac{1}{2}$ chance of transitioning to Committed Active $(C^A_i)$ and a $\frac{1}{2}\%$ chance of Committed Nest $C^N_i$ after having detected or been notified of a quorum. 

$C^N_i$ agents head to the home nest to inform others of the move, while $C^A_i$ agents randomly wander the grid to find stragglers. Agents in quorum states continue to wander until they have sensed quorum for $t_Q$ time steps, whereupon they return to the new selected site $s_i$. 

The resulting agent state set $SR$ is a product of the $6$ core states needed in the state model as well as a number of auxiliary variables such as an agent's destination, the names of the sites it favors or has sensed quorum for, and parameters for an agent's random walk when exploring the grid.

Since in the house hunting problem (unlike other problems like task allocation), an agent never modifies the environment, an agent's proposed new vertex state is always the same as the old vertex state. Therefore, phase two of $\delta$ is not needed to reconcile conflicting vertex state suggestions from agents. 

The transition function $\alpha$, which for each agent returns a proposed new vertex state $sv'$, agent state $sr'$ and direction of motion works as follows. The agent never modifies the grid, so $sv'=sv$. The agent state $sr'$ and direction $d$ are calculated according to the core transitions and the auxiliary variables needed to keep track of those transitions. For example, when an agent is headed towards a site, the direction $d$ is calculated to try to match a straight-line traversal to the site from the agent's starting location as close as possible. When an agent is staying within a site, the direction $d$ is calculated to be a random walk within the site boundaries. 

\sloppy
The total set of variables parameters is $\{P_A, P_N, D, t_Q, t_\beta, q_{MIN}, q_{MAX}\}$, as well as the site locations $(x_i^1, y_i^1), (x_i^2, y_i^2)$ and quality $s_i.q$. In Section \ref{sec:results}, we explore how changes in $q_{MIN}$, $q_{MAX}$, and the site locations and quality impact the accuracy, decision time, and split decisions made by the model. 

\section{Results}\label{sec:results}
The model was tested in simulation using Pygame, with each grid square representing 1cm$^2$. Agents moved at 1 cm/s, with one round representing one second. We chose this speed because even the lowest cost robots are still able to move at 1cm/s \cite{rubenstein2012kilobot}. Agents had an influence radius of $2$. All simulations were run using 100 agents, and a messaging rate of $1/15$. We let the abandonment timeout $t_\beta = \frac{5}{L}$ and the quorum timeout $t_Q = \frac{1}{L}$.

For each set of trials, we evaluated accuracy (measured as the fraction of agents who chose the highest quality nest), decision time (measured as the time it took for all agents to arrive at the nest they committed to), and split decisions (measured as the number of trials where not all agents committed to the same nest). 

\subsection{Further Nest of Higher Quality}
House hunting ants are capable of choosing further, higher quality sites over closer, lower quality ones \cite{franks2008can}. When the far site and the near site are of equal value, ants consistently choose the closer one. To test our model's ability to produce the same behavior, we replicated the experimental setups in \cite{franks2008can}.

Three different distance comparisons were tested, with a further, higher quality nest of quality $0.9$ being 2x, 3x, and 9x as far as a lower quality nest of quality $0.3$ on the path from the high quality nest to the home nest. We included a control setup for each of these distance comparisons where both the far and close nest were quality $0.3$. The arena size was $N=16, M=80$ for the 2x case, $N=18, M=180$ for the 3x case, and $N=18$, $M=300$ for the 9x case.

We tested our model using two different quorum parameters. In one test, we had $q_{MIN} = q_{MAX} = 4$, intended to represent the behavior of previous models with a fixed quorum threshold. In the other setup, $q_{MIN} = 4$ and $q_{MAX} = 7$, allowing our model to use the new feature of scaling the quorum threshold with site quality. We ran 100 trials for each set of parameters. 

\begin{figure}
    \centering
    \includegraphics[width=0.45\textwidth]{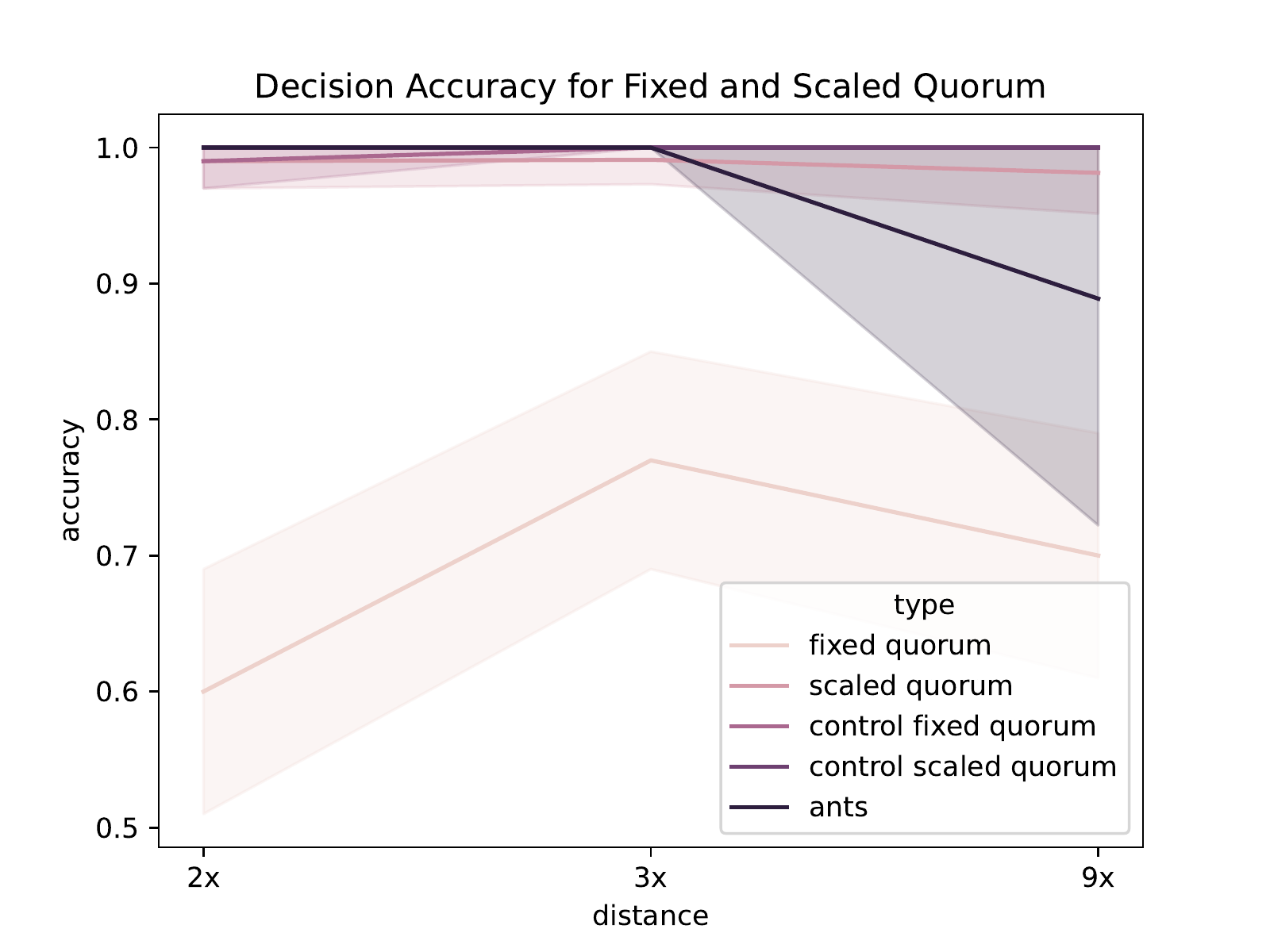}
    \includegraphics[width=0.45\textwidth]{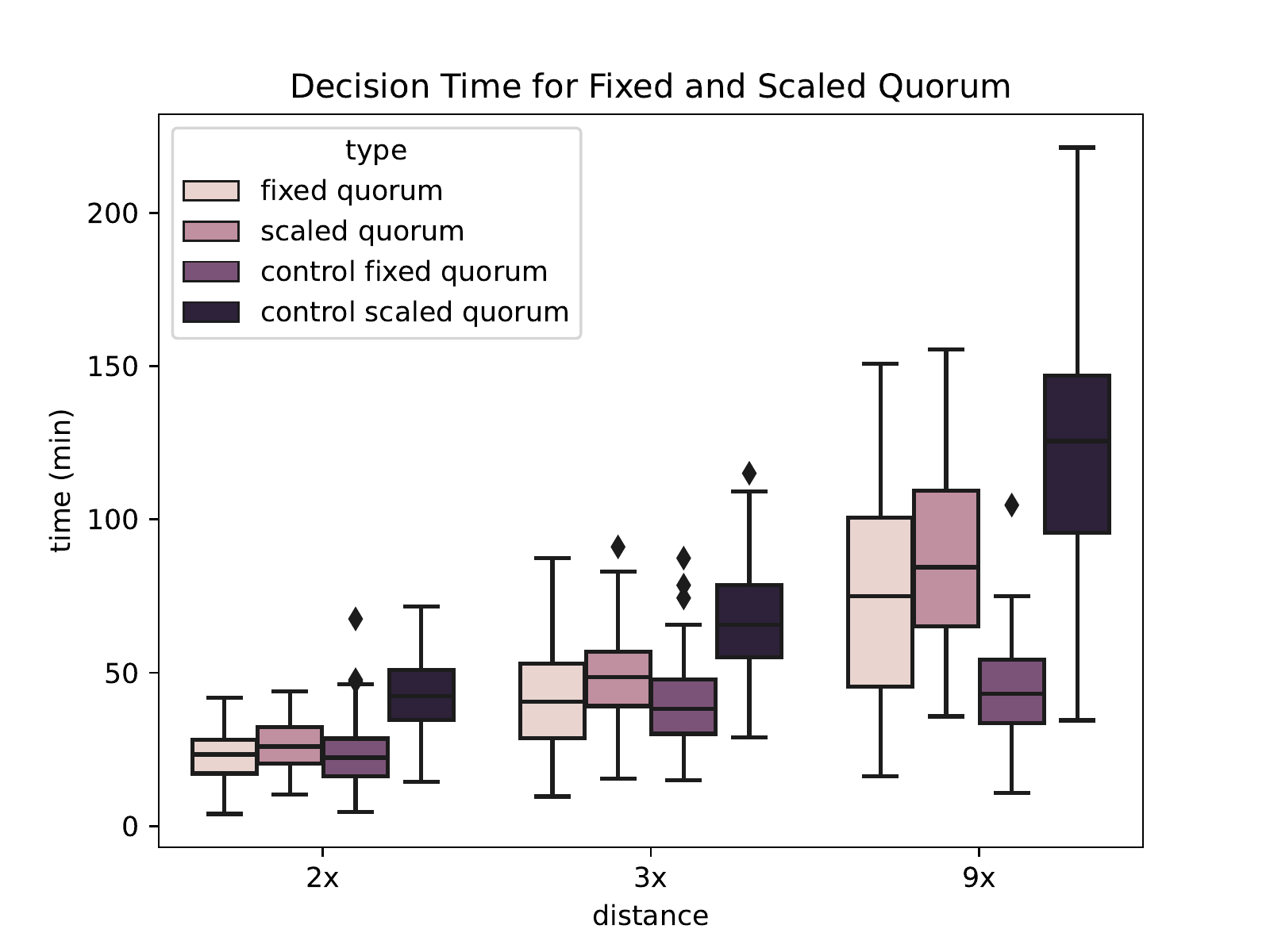}
    \caption{Decision Time and Accuracy for far nests 2, 3, and 9 times as far from the home nest. Fixed quorum indicates the fixed threshold value of $4$, and scaled quorum indicates $q_{MIN}=4, q_{MAX}=7$. The accuracy for the actual ants is taken from \cite{franks2008can}.}
    \label{fig:franks_results}
    \vspace{-0.5cm}
\end{figure}

As seen in Figure \ref{fig:franks_results}, using a scaled threshold significantly improved accuracy from using a fixed one. In the control case, both the fixed and scaled quorum threshold achieved high accuracy, with all accuracies being greater than $99\%$. In cases where the far site was of higher quality, the decision time for fixed and scaled quorum was comparable. However, the scaled quorum threshold took significantly (Welch's T-test, p=0.05) more time in the control case to decide.

Furthermore, as seen in Figure \ref{fig:franks_results}, our model successfully chose the further site with comparable (or significantly higher in the 9x case) accuracy than ants themselves, indicating that our model is on par with the ants.

\subsection{Effects of Lower Quality Nest Being In the Way}
To isolate the effects of the low quality nest being in the way of the high quality nest, we tested our model where the high quality nest (quality $0.9$) was one of $\{2, 3, 4, 5, 6, 7, 8, 9\}$ times further than the low quality nest (quality $0.3$), but in opposite directions of the home nest. We compared model performance when the low quality nest was in the way of the home nest. We ran tests with $N=18$, $M=300$, with the low quality nest always $30$cm from home. We again tested a fixed ( $q_{MIN}=q_{MAX}=4$) and  scaled($q_{MIN}=4, q_{MAX}=7$) quorum threshold on these setups. $100$ trials were conducted for each set of parameters.

\begin{figure}
    \vspace{-0.4cm}
    \centering
    \includegraphics[width=0.45\textwidth]{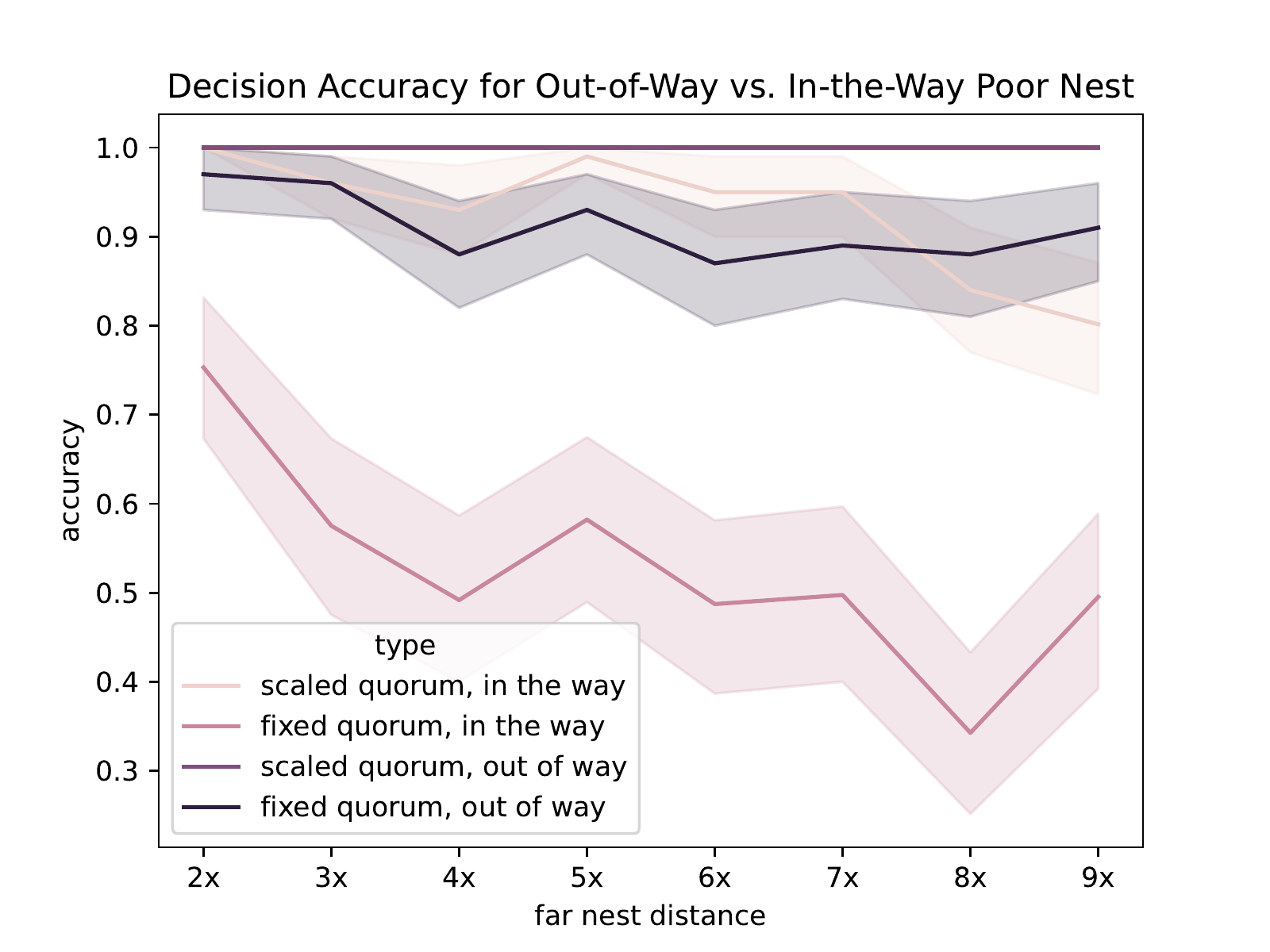}
    \includegraphics[width=0.45\textwidth]{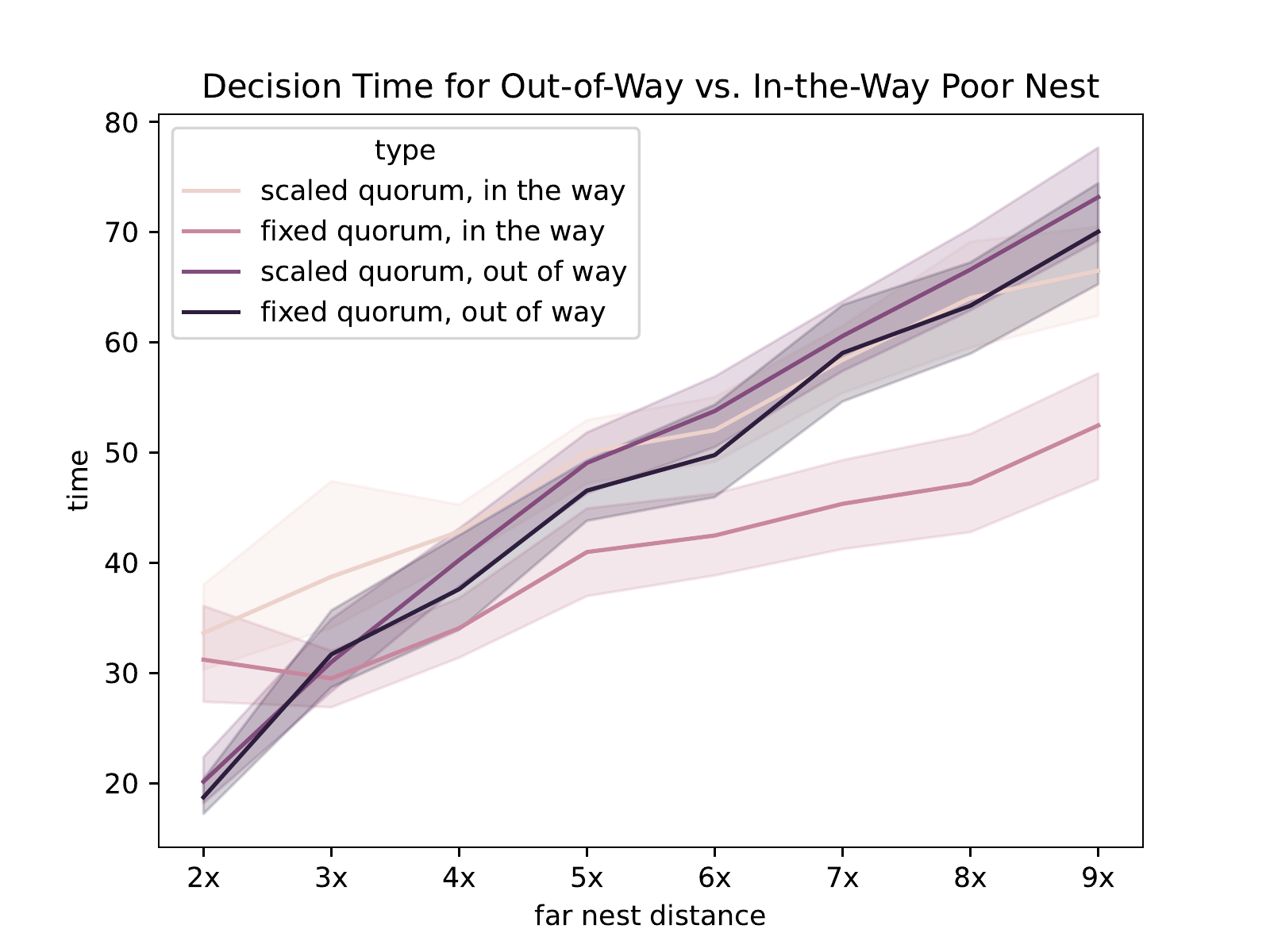}
    \caption{Decision Time and Accuracy for far nests $2-9$ times further than the close nest for both fixed and scaled quorums. In the in-the-way setup, the home nest, low quality nest, and high quality nest were lined up in that order. In the out of way setup, the low quality nest, home nest, and high quality nest were lined up in that order.}
    \label{fig:out_vs_in}
    
\end{figure}

Figure \ref{fig:out_vs_in} shows that for the out-of-way setup, the scaled quorum performs significantly (Welch's T-test, p=0.05) more accurately than the fixed quorum on all far nest distances. For the in-the-way setup, the scaled quorum performs significantly better (Welch's T-test, p=0.05) when the far nest is $3$x further or more. Note it is harder for the fixed quorum to solve the in-the-way problem accurately compared to the out-of-way problem (Welch's T-test, p=0.05). It is likewise harder for the scaled quorum to solve the in-the-way problem when the far nest is $\{3, 4, 6, 7, 8, 9\}$ times further (Welch's T-test, p=0.05), showing that the in-the-way problem is harder to solve for site selection algorithms.

For distances $3$x or further, there is no significant difference between the decision times for the fixed out-of-way, scaled out-of-way, and scaled in-the-way setups. For distances $5$x and further, the fixed quorum takes significantly less time than the other setups but suffers in decision accuracy (Welch's T-test, p=0.05) compared to the other three setups. 
\subsection{Effects of Magnitude of Difference in Site Quality}
Because site quality affects the quorum threshold, we expect it to be harder for agents to correctly choose a high quality far site when it is only slightly better than than nearby lower quality sites. This is because the difference in quorum threshold is less pronounced for sites of similar quality. For two equidistant nests, the algorithm should consistently choose the best site as it has in past work, so the absolute difference in site quality should not matter.

\begin{figure}
    \vspace{-0.4cm}
    \centering
    \includegraphics[width=0.45\textwidth]{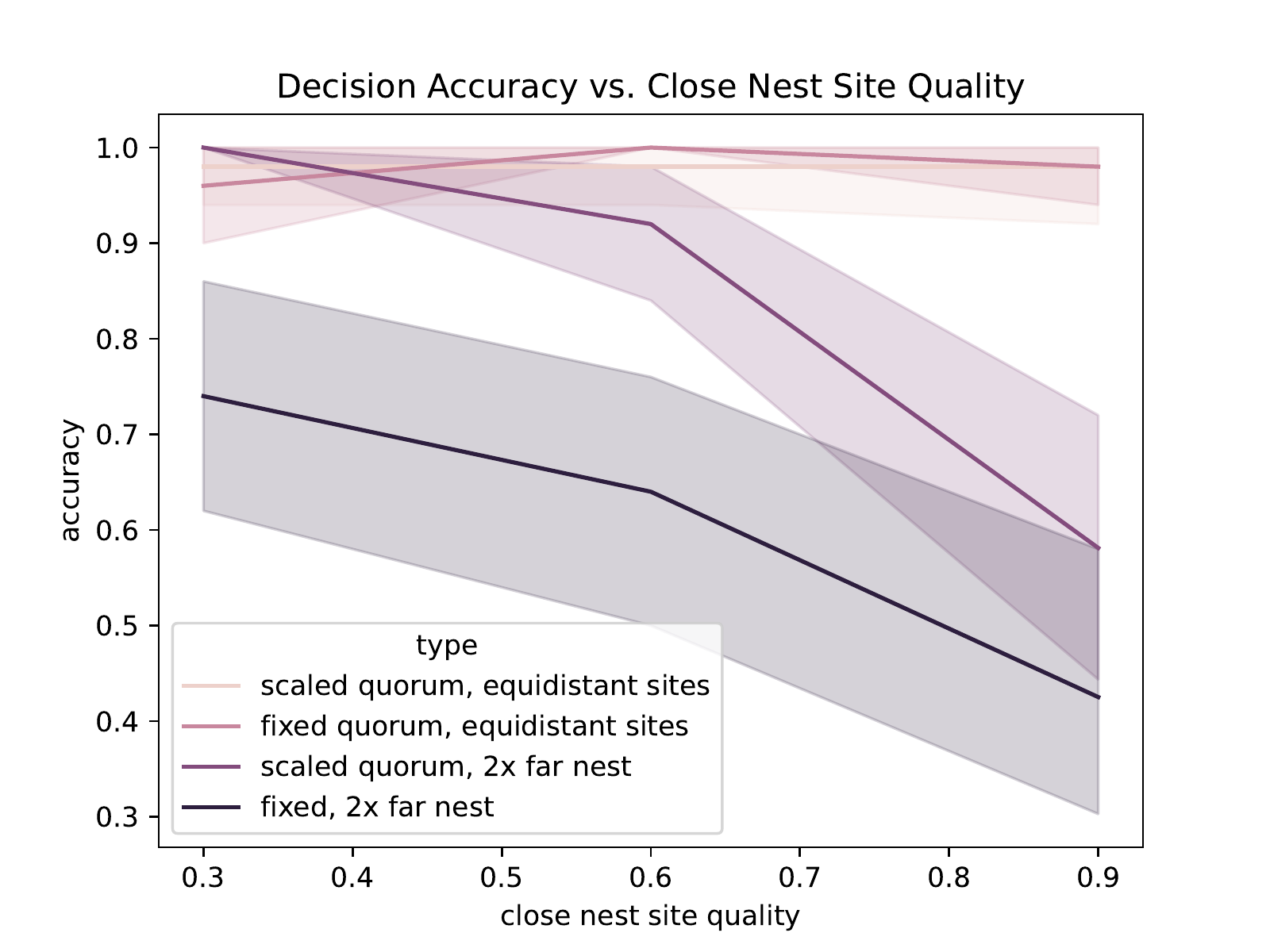}
    \includegraphics[width=0.45\textwidth]{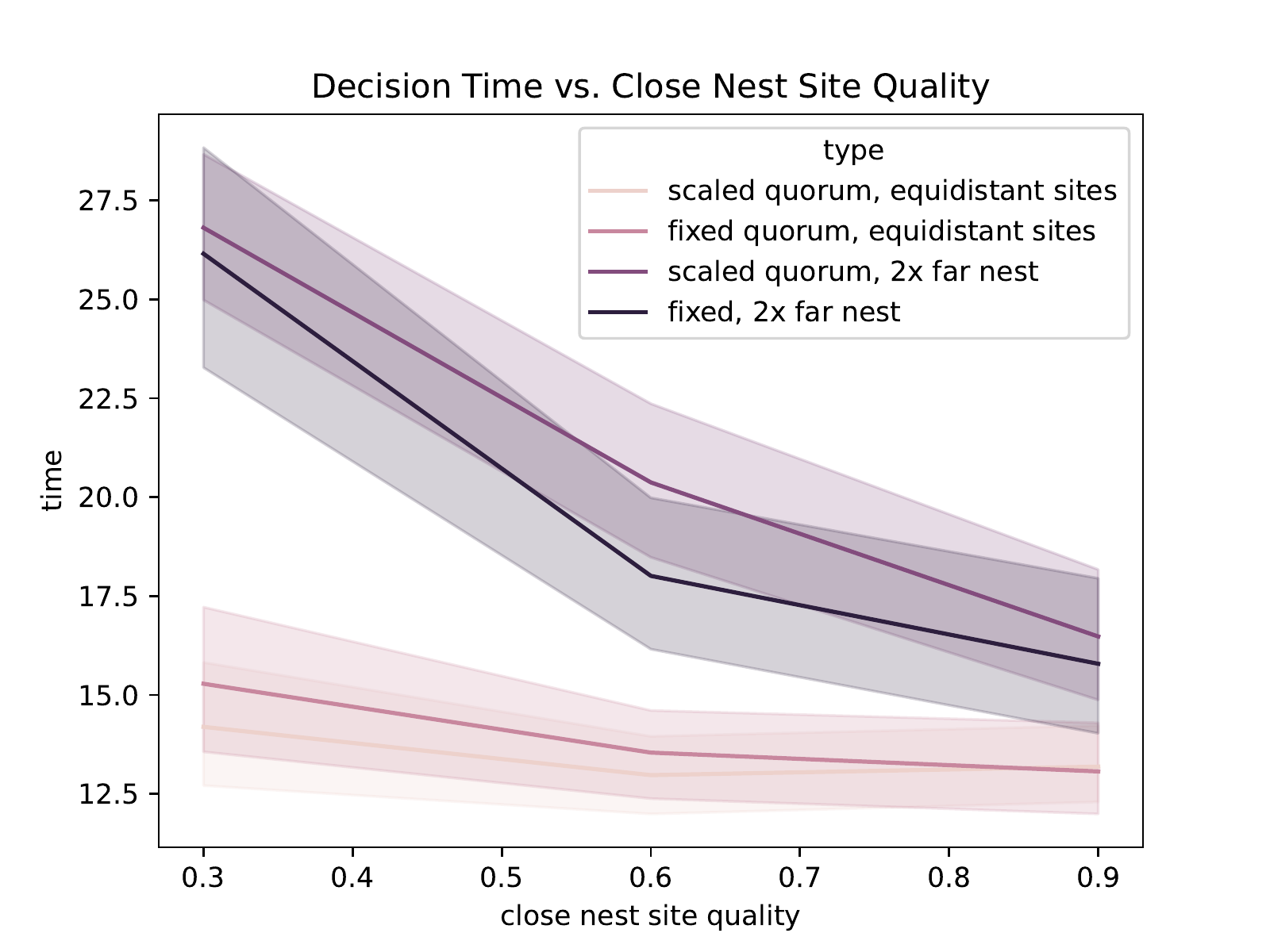}
    \caption{Decision Accuracy and Time given varying differences in site quality between the near and the far nest.}
    \label{fig:site_vals}
\end{figure}

To test these effects, we used the setup in Section 4.1 where the further nest was $2$x (60 cm) as far as the in-the-way close nest (30 cm), and compared it to an equidistant setup where both candidate nests were $30$ cm away from the home nest in opposite directions. We tested both a fixed quorum $q_{MAX}=q_{MIN}=4$ and a scaled quorum on these setups $q_{MAX}=7, q_{MIN}=4$. We varied the quality of the near nest in the set of potential values $\{0.3, 0.6, 0.9\}$, corresponding to quorum thresholds of $\{6, 5, 4\}$ respectively, with the far nest having quality $1.0$. (In the equidistant case, we varied the quality of one nest while the other had quality $1.0$.) Figure \ref{fig:site_vals} shows the resulting accuracy and decision time.

As predicted, a smaller difference in site quality / quorum threshold led to significantly (Welch's T-test, p=0.05) lower decision accuracy for the non-equidistant setup. In the equidistant setup, agents were able to achieve a near-100\% outcome regardless of magnitude of differences in site quality. However, in the unbalanced setup, we confirmed that for larger differences in site quality, the algorithm comes to a more accurate decision, showing that non-equidistant candidate nest setups cause sensitivity to absolute site value differences that can't be seen in the equidistant setup.

\section{Discussion}\label{sec:discussion}
The results demonstrate our model's ability to improve accuracy when choosing from a higher quality, further site and a lower quality, closer site. This improvement comes at the cost of a higher decision time when converging on a lower quality site, because the quorum threshold for low quality sites is higher in our model. This higher decision time is reasonable and represents hesitance when committing to a poor quality option in the hopes of finding a better one.

Our model also demonstrated the extra difficulty that comes with a lower quality site being in the path from the home nest to a high quality site. Qualitative observation showed that agents travelling back and forth between the far site and the home nest often unintentionally contributed to a quorum in the poor quality, in-the-way site as they travelled through it.  We showed that using a scaled quorum threshold as opposed to a fixed one is an effective way of significantly increasing decision accuracy.
However, even if the closer, poor quality site is completely out of the way of the far, high quality site, Figure \ref{fig:out_vs_in} shows that using a scaled quorum can still help to improve accuracy.

Figure \ref{fig:site_vals} shows that our model is still successful when candidate sites are equidistant from home, as is most commonly tested. We also show that an equidistant setup is not influenced by the absolute difference between candidate site qualities. Contrarily, in the setup with a further, high quality nest, it is harder to make an accurate decision the smaller the quality difference between the high and low quality nests. Note that it is also less grievous of an error to choose the low quality nest when the quality difference is small.

We observed a shorter decision time in conjunction with lower accuracy, similar to the time-accuracy trade off in natural swarms \cite{chittka2003bees,heitz2014speed}. In each set of 100 trials run on our setups had at most $2$ split decisions, indicating our model is successful in keeping the swarm together even when migrating to the further nest. 

\section{Future Work}\label{sec:future}
While our model has made strides in being more accurate when choosing between sites with different geographical distributions, many site setups have yet to be tested. Future work could introduce obstacles to the environment, or try to adapt the house hunting model to an arena with continuous site values. 

Our model suggests that a quorum threshold that scales with site quality leads to more accurate site selection. Future work could explore if actual ants do the same and use this information to create more accurate models.

Lastly, while our model is hard to analyze without making simplifications (because it involves agents physically moving in space), future work could try to develop analytical bounds. One method we envision is simplifying the chance of each site being discovered to a fixed probability and trying to model agent population flow between the different model states, similar to \cite{reina-bee-model}, which does this for candidate sites all with an equal chance of discovery. 

\bibliographystyle{splncs04}
\bibliography{mybibliography}

\clearpage\section*{Appendix A: Formal General Model}
Below we present an alternative, more formal description of our general model from section 3.1. 

\subsection{General Model}
We assume a finite set $R$ of agents, all of which have a state set $SR$ of potential states. Agents move on a discrete rectangular grid of size $n \times m$, formally modelled as directed graph $G = (V,E)$ with $|V| = mn$. Edges are  bidirectional, and we also include a self-loop at each vertex. Vertices are indexed as $(x, y)$, where $0 \leq x, y \leq n-1$. Each vertex also has a state set $SV$ of potential states.

\subsection{Grid and Agent Setup}
We define four kinds of configurations, global vs. local, and ordinary vs. transitory.  The transitory configurations are used as intermediate steps in defining system executions.  

\paragraph{(Global) configurations:}

A \emph{(global) configuration} $C$ is a triple of mappings, $(svmap, srmap, locmap)$, where:
\begin{itemize}
    \item 
    $svmap: V \rightarrow SV$ is the vertex state mapping, which assigns a vertex-state to each vertex,
    \item
    $srmap: R \rightarrow SR$ is the agent state mapping, which assigns an agent-state to each agent, and
    \item 
    $locmap: R \rightarrow V$ is the location mapping, which assigns a location to each agent.
\end{itemize}

\paragraph{Local configurations:}
A \emph{local configuration} $C'$ is intended to capture the contents of one vertex/square).  It is a triple $(sv, myagents, srmap)$, where:
\begin{itemize}
    \item 
    $sv \in SQ$ is the vertex-state of the given vertex,
    \item
    $myagents \subseteq R$ is the set of agents at the vertex, and
    \item 
    $srmap: myagents \rightarrow SR$ assigns an agent-state to each agent at the vertex.
\end{itemize}

Define an operator $project(C,v)$, which takes a global configuration $C$ and projects it to give a local configuration of vertex $v$.
Namely, if $C = (svmap, srmap, locmap)$, and $v$ is any vertex, then $project(C,v)$ is the local configuration
$(svmap(v), R', srmap \lceil R')$, where $R' = \{r \in R : locmap(r) = v\}$.
That is, we pick out the state of the given vertex $v$, and the states for all the agents located at $v$.

We also define an inverse operation called $merge$, which combines local configurations to yield a global configuration.
We assume a collection of "compatible" local configurations 
$\{C'(v) : v \in V \}$, where "compatible" here means that the unique vertices below exist.
The resulting global configuration $C$ is defined by:
\begin{itemize}
    \item 
    For every $v \in V$, $svmap(v) = C'(v).sv$, that is, take the state of $v$ from the local configuration belonging to vertex $v$.
    \item
    For every $r \in R$, $locmap(r)$ is the unique vertex $v$ such that $r \in C'(v).myagents$, and $srmap(r) = C'(locmap(r)).srmap(r)$.
\end{itemize}

\subsection{Transitory Configurations}
We also have a notion of transitory configuration, which is used as an intermediate stage between two ordinary configurations, in constructing executions.  It represents agents in motion from one vertex to another. 

A \emph{transitory configuration} $T$ is a triple of mappings $(svmap, srmap, edgemap)$, where
\begin{itemize}
\item
$svmap$ and $srmap$ have the same types as for ordinary configurations, and
\item
$edgemap: R \rightarrow E$ assigns a directed edge to each agent.
\end{itemize}

This represents newly-computed states for all vertex and agents, plus the edges on which the agents are traveling to get to their next locations.

\paragraph{Local transitory configurations:}
A \emph{local transitory configuration} $T'$ is a quadruple $(sv, myagents, srmap, dirmap)$, where
\begin{itemize}
\item 
$sv$, $myagents$, and $srmap$ are as in the definition of a local configuration, and
\item
$dirmap: myagents \rightarrow \{R,L,U,D,S\}$ is the direction mapping, which assigns a travel direction to each agent.  Direction $S$ means to stay at the vertex.
\end{itemize}
This represents newly-computed states for a single vertex and its agents, plus directions of travel for the local agents.

As before, define an operator $project(T,v)$, which takes a global transitory configuration $T$ and projects it to give a local transitory configuration $T(v)$ for vertex $v$.
Namely, if $T = (svmap, srmap, edgemap)$, and $v$ is any vertex, then $project(T,v)$ is the local transitory configuration
$(svmap(v), myagents, srmap \lceil myagents, dirmap)$,
where:
\begin{itemize}
\item
$myagents = \{r \in R : edgemap(r) \mbox{ is an edge with } source = v \}$, and
\item
For each $r \in myagents, dirmap(r)$ is the direction of the edge $edgemap(r)$, that is, $R$ if the edge goes from $v$ to its right neighbor, etc.
\end{itemize}

We also define an inverse, $merge$ operation, which combines local transitory configurations to give a global transitory configuration.
We assume a collection of "compatible" local configurations $\{T'(v) : v \in V\}$, where "compatible" means that the unique vertices below exist.  
The resulting global transitory configuration $T$ is defined by:
\begin{itemize}
    \item 
    For every $v \in V$, $svmap(v) = T'(v).sv$,
\item
For every $r \in R$, let $v$ be the unique vertex such that $r \in T'(v).myagents$; then define:
\begin{itemize}
    \item $srmap(r) = T'(v).srmap(r)$.
    \item $edgemap(r)$ is the directed edge from $v$ to $w$, where $w$ is the vertex that is the target of the edge from $v$ in the direction given by $T'(v).dirmap(r)$.
\end{itemize}
\end{itemize}

Given a transitory configuration $T = (svmap, srmap, edgemap)$, we can map it to an ordinary configuration using $C = move(T)$ , which is the same as $T$ except that, instead of edgemap, we have locmap, where for each $r \in R$, $locmap(r) = target(edgemap(r))$.

\subsection{Local transitions}
The transition of a vertex $v$ may be influenced by the local configurations of nearby vertices in addition to itself. We define an \textbf{influence radius} $I$, which is the same for all vertices, to mean that vertex indexed at $(x,y)$ is influenced by all vertices $\{(a,b) | a \in [x-I, x+I], b\in [y-I, y+I]\}$, where $a$ and $b$ are integers mod $n$. 
We can use this influence radius to create a local mapping $M_v$ from local coordinates to the neighboring local configurations. Thus, for a vertex $v$ at location $(x,y)$, we produce $M_v$ such that $M_v(a,b) \rightarrow C'(w)$ where $w$ is the vertex located at $(x+a, y+b)$ and $ -I < a,b < I$. This influence radius is representative of a sensing and communication radius for agents.

We have a local transition function $\delta$, which maps all the information associated with one vertex and its influence radius at one time to new information that can be associated with the vertex at the following time.  It also produces directions of motion for all the agents at the vertex.

Formally, for a vertex $v$, $\delta$ maps the mapping $M_v$ to a probability distribution on local transitory configurations of the form $(sv_1, myagents, srmap_1, dirmap_1)$, where:
\begin{itemize}
\item
$sv_1 \in SV$ is the new state of the vertex,
\item
$srmap_1: myagents \rightarrow SR$ is the new agent state mapping, for agents currently at the vertex, and
\item
$dirmap_1: myagents \rightarrow \{R,L,U,D,S\}$ gives directions of motion for all the agents currently at the vertex.
\end{itemize}

\subsection{Local transition function $\delta$}
The local transition function $\delta$ is further broken down into two phases as follows.
\paragraph{Phase One:} Each agent in vertex $v$ uses the same probabilistic transition function $\alpha$, which maps the agent's state $sr \in SR$, location $(x,y)$, the vertex state of the location $sv \in SV$, and the mapping $M_v$ to a distribution over new suggested vertex state $sv'$, agent state $sr'$, and direction of motion $d \in \{R, L, U, D, S\}$.
\paragraph{Phase Two:} A rule $L$ is used to reconcile the different vertex states suggested by each agent at the vertex and select one final vertex state. The rule also determines for each agent whether they may transition to state $sr'$ and direction of motion $d$ or stay at the same location with original state $sr$.

\subsection{Probabilistic execution}

The system executes probabilistically, moving through an infinite sequence of configurations $C_0, C_1, C_2,\ldots,$ each derived probabilistically from the previous one.

We first describe how the system moves from any configuration $C_t$ to the next configuration $C_{t+1}$.
This is a two-phase process where the first phase is probabilistic and the second is deterministic.
The first phase yields a transitory configuration, and the second phase converts that to an ordinary configuration.

\paragraph{Phase 1:}
For each vertex $v$ independently, use $project(C_t,w)\ \forall w \in IR(v)$ to obtain the local mapping $M_v$.
Then apply $\delta$ to $M_v$, and sample the resulting distribution to select a local transitory configuration for $v$.  
Apply $merge$ to combine these independent results for all $v$ as described above, to obtain a global transitory configuration.

\paragraph{Phase 2:}
For the transitory configuration $T$ resulting from Phase 1, compute $move(T)$ to get the new configuration.
\end{document}